# Variational perturbation expansion for strong-coupling coefficients of the anharmonic oscillator

W. Janke[1,2] and H. Kleinert[2]

[1] *Institut für Physik, Johannes Gutenberg-Universität Mainz, Staudinger Weg 7, 55099 Mainz*
[2] *Institut für Theoretische Physik, Freie Universität Berlin, Arnimallee 14, 14195 Berlin*

As an application of a recently developed variational perturbation theory we find the first 22 terms of the convergent strong-coupling series expansion for the ground state energy of the quartic anharmonic oscillator.

Variational perturbation theory yields uniformly and exponentially fast converging expansions for any quantum mechanical system [1–3]. In this note we adapt this theory the problem of calculating strong-coupling expansions. As an example, we consider the quantum mechanical anharmonic oscillator and find the expansion coefficients with a high degree of accuracy. The input is provided by the exact Rayleigh-Schrödinger perturbation coefficients of the ground-state energy up to order 251, obtained from the recursion relations of Bender and Wu [4]. The resulting strong-coupling coefficients will be calculated up to the 22nd order with a precision of about 20 digits.

The potential of the anharmonic oscillator is

$$V(x) = \frac{\omega^2}{2}x^2 + \frac{g}{4}x^4 \qquad (\omega^2, g > 0). \tag{1}$$

The Rayleigh-Schrödinger perturbation theory yields a power series expansion

$$E(g) = \omega \sum_{l=0}^{\infty} e_l^{\text{BW}} \left(\frac{g/4}{\omega^3}\right)^l \tag{2}$$

where $e_l^{\text{BW}}$ are rational numbers

$$\frac{1}{2}, \frac{3}{4}, -\frac{21}{8}, \frac{333}{16}, -\frac{30885}{128}, \ldots. \tag{3}$$

The superscript BW refers to Bender and Wu. Using the symbolic algebra program MAPLE we generate the first 251 terms exactly.

The series (2) cannot be used for an evaluation of the energy—it has a zero radius of convergence due to the factorial growth of the coefficients $e_l^{\text{BW}}$. Only for small couplings $g < 0.1$, it yields reasonable approximations if truncated at a finite order $N$, optimally at the integer closest to $3/4g$.

The recently developed variational perturbation theory [2,3] converts the divergent series (2) into an exponentially fast convergent one. The procedure goes as follows (see Section 5.13 of Refs. [2,3]). First, the harmonic term of the potential is split into a new harmonic term with a trial frequency $\Omega$, and a remainder:

$$\frac{\omega^2}{2}x^2 = \frac{\Omega^2}{2}x^2 + \left(\frac{\omega^2}{2} - \frac{\Omega^2}{2}\right)x^2. \tag{4}$$

After rewriting

$$V(x) = \frac{\Omega^2}{2}x^2 + V_{\text{int}}(x), \tag{5}$$

with an interaction

$$V_{\text{int}}(x) = \frac{g}{4}(rx^2 + x^4), \quad r = \frac{2}{g}(\omega^2 - \Omega^2), \tag{6}$$

we perform a perturbation expansion in powers of $g$ at a fixed $r$,

$$E_N(g, r) = \Omega \sum_{l=0}^{N} e_l(r) \left(\frac{g/4}{\Omega^3}\right)^l. \tag{7}$$

The calculation of the new series up to a specific order $k$ requires only littel additional work, being easily obtained from the ordinary perturbation series (2) by replacing $\omega$ by $\sqrt{\Omega^2 + gr/2}$, and by reexpanding (2) in powers of $g$ up to the $N$th order. This yields the reexpansion coefficients

$$e_l(r) = \sum_{j=0}^{l} e_j^{\text{BW}} \binom{(1-3j)/2}{l-j} (2r\Omega)^{l-j}. \tag{8}$$

The truncated power series

$$W_N(g, \Omega) := E_N\left(g, \frac{2}{g}(\omega^2 - \Omega^2)\right) \tag{9}$$

is certainly independent of $\Omega$ in the limit $k \to \infty$. At any finite order, however, it *does* depend on $\Omega$, the approximation having its fastest speed of convergence where it depends least on $\Omega$. If we denote the order-dependent optimal value of $\Omega$ by $\Omega_N$, the quantity $W_N(g, \Omega_N)$ is the new approximation to $E(g)$ [5]. In terms of the dimensionless constants

$$\hat{g} = g/\Omega^3, \quad \hat{\omega} = \omega/\Omega, \tag{10}$$

the approximation can be written as

$$W_N = (g/\hat{g})^{1/3} w_N(\hat{g}, \hat{\omega}^2). \tag{11}$$

From the approximate energies (9) it is easy to derive simple formulas for the coefficients of the strong-coupling expansion [3]. We simply expand the function $w_N(\hat{g}, \hat{\omega}^2)$ in powers of $\hat{\omega}^2 = (g/\omega^3)^{-2/3} \hat{g}^{2/3}$ and find

$$W_N = (g/4)^{1/3} \left[ \alpha_0 + \alpha_1 \left( \frac{g/4}{\omega^3} \right)^{-2/3} + \alpha_2 \left( \frac{g/4}{\omega^3} \right)^{-4/3} + \ldots \right], \quad (12)$$

with the coefficients

$$\alpha_n = \frac{1}{n!} \hat{g}^{(2n-1)/3} w_N^{(n)}(\hat{g}, 0). \quad (13)$$

Here $w_N^{(n)}(\hat{g}, 0)$ denotes the $n$'th derivative of $w_N(\hat{g}, \hat{\omega}^2)$ with respect to $\hat{\omega}^2$ at $\hat{\omega}^2 = 0$. To calculate these derivatives, we note that $\hat{\omega}^2$ enters the reexpansion coefficients (8) in the form

$$e_l = \sum_{j=0}^{l} e_j^{BW} \binom{(1-3j)/2}{l-j} (4(\hat{\omega}^2 - 1)/\hat{g})^{l-j}. \quad (14)$$

The quantities $w_N^{(n)}(\hat{g}, 0)$ are therefore given by

$$w_N^{(n)}(\hat{g}, 0) = \left( \frac{d}{d\hat{\omega}^2} \right)^n e_l |_{\hat{\omega}=0} = \sum_{j=0}^{l} e_j^{BW} \binom{(1-3j)/2}{l-j}$$
$$\times \frac{(-1)^n}{n!} \left( \prod_{m=0}^{n-1} (l-j-m) \right) (-4/\hat{g})^{l-j}. \quad (15)$$

The optimal value of $\Omega_N$ has the $N$-dependence (see Ref. [3])

$$\Omega_N^3 = gcN \left( 1 + 6.85/N^{2/3} \right) \quad (16)$$

where the coefficient $c$ is the solution of a simple transcendental equation ($c = 0.186\,047\,272\,987\,397\,512\,984\,554\,740\,462$. eyeball fit to extremal values of $\Omega$. The corresponding order-dependent values of the dimensionless coupling constant $\hat{g} \equiv g/\Omega_N^3$ are inserted into (13) and produce the coefficients shown in Table I. Our result for $\alpha_0$ agrees to all 23 digits with the most accurate value for $\alpha_0$ available in the literature [6]:

$$\alpha_0 = 0.667\,986\,259\,155\,777\,108\,270\,962\,016\,919\,860$$
$$199\,430\,404\,936\,984\,060\,455\,976\,663\,80. \quad (17)$$

For $\alpha_1$ and $\alpha_2$, our results are consistent with but considerably more accurate than previous results in Ref. [7] and Refs. [8,9] [$\alpha_1 = 0.143\,668\,800$, $\alpha_2 = -0.008\,627\,58$].

As a further check we have evaluated our strong-coupling series at $g/4 = 0.1, 0.3, 0.5, 1, 2$ (setting $\omega = 1$) and compared the numbers with the very precise lower and upper bounds of Vinette and Čížek [6]. Table II shows that the energies are accurate to about 20 digits for all couplings $g/4 \geq 1$. The accuracy is limited by the precision of the $\alpha_n$.

Note that our strong-coupling expansion gives good results down to very small couplings $g$—even at $g/4 = 0.1$, the result agreeing to 7 digits with the bounds in [6]. If one wants to work at such small couplings, one can easily calculate higher coefficients $\alpha_n$ to increase the accuracy.

In Fig. 1 we show the approach of $\alpha_n$ to the asymptotic value given in Table I by plotting

$$\Delta_N \equiv |(\alpha_n)_N - \alpha_n| \quad (18)$$

on a logarithmic scale. The periodic structure in the data is caused by an oscillatory approach of $(\alpha_n)_N - \alpha_n$ to zero (see Fig. 2). The envelope of these oscillations follows the curve

$$\Delta_N = \exp\left(-\kappa_0 - \kappa_1 N^{1/3}\right). \quad (19)$$

For the leading term $\alpha_0$ of the strong-coupling expansion, the behavior (16) suggests a coefficient $\kappa_1 \approx 9.7$ [3]. From the eyeball fits shown in Fig. 1 we extract $\kappa_1 \approx 9.42, 9.05, 8.05,$ and $7.18$ for $\alpha_n$ with $n = 0, 1, 5$ and 10, respectively. These estimates should be taken with some care, however, since we do not know how close we are to the asymptotic regime for $N \approx 65 \ldots 251$, where the natural expansion parmeter is $1/N^{1/3} \approx 0.25 \ldots 0.16$. The $N$-dependence of the coefficient $\alpha_0$, for instance, is just as well fitted by the dashed line with $\kappa_1 = 9.7$.

We conclude by mentioning that variational perturbation theory allows also for a calculation of the imaginary parts of energies on the left-hand cut in the complex coupling constant plane [10]. The results not only improve the semiclassical values [11] in the tunneling regime, but also determines the imaginary part at large negative $g$ [10,12], where the decay proceeds by sliding rather than tunneling.

W.J. thanks the Deutsche Forschungsgemeinschaft for a Heisenberg fellowship.

TABLE I. Strong-coupling perturbation coefficients $\alpha_n$.

| $n$ | $\alpha_n$ |
| --- | --- |
| 0 | 0.667 986 259 155 777 108 270 96 |
| 1 | 0.143 668 783 380 864 910 020 3 |
| 2 | −0.008 627 565 680 802 279 128 |
| 3 | 0.000 818 208 905 756 349 543 |
| 4 | −0.000 082 429 217 130 077 221 |
| 5 | 0.000 008 069 494 235 040 966 |
| 6 | −0.000 000 727 977 005 945 775 |
| 7 | 0.000 000 056 145 997 222 354 |
| 8 | −0.000 000 002 949 562 732 712 |
| 9 | −0.000 000 000 064 215 331 954 |
| 10 | 0.000 000 000 048 214 263 787 |
| 11 | −0.000 000 000 008 940 319 867 |
| 12 | 0.000 000 000 001 205 637 215 |
| 13 | −0.000 000 000 000 130 347 650 |
| 14 | 0.000 000 000 000 010 760 089 |
| 15 | $-0.000\,000\,000\,000\,000\,445\,890\,1$ |
| 16 | $-0.000\,000\,000\,000\,000\,058\,989\,8$ |
| 17 | $0.000\,000\,000\,000\,000\,019\,196\,00$ |
| 18 | $-0.000\,000\,000\,000\,000\,003\,288\,13$ |
| 19 | $0.000\,000\,000\,000\,000\,000\,429\,62$ |
| 20 | $-0.000\,000\,000\,000\,000\,000\,044\,438$ |
| 21 | $0.000\,000\,000\,000\,000\,000\,003\,230\,5$ |
| 22 | $-0.000\,000\,000\,000\,000\,000\,000\,031\,4$ |

The solid straight lines are the best eyeball fits to the envelope of the data. The dashed line in the $\alpha_0$-data has a slope equal to the theoretically expected value 9.7 [3] .

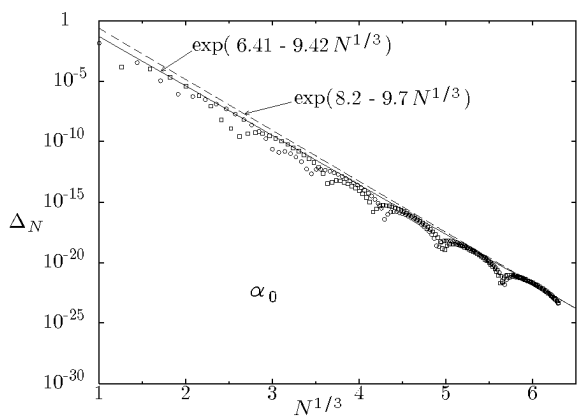

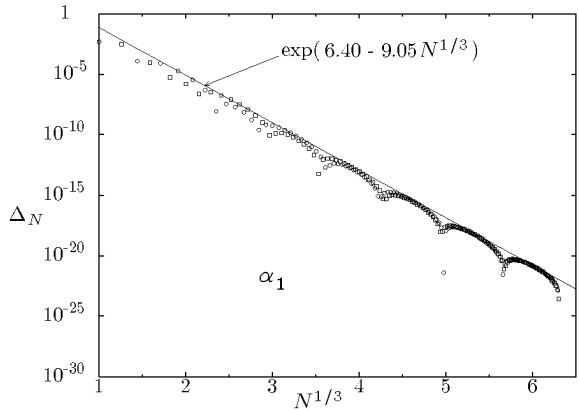

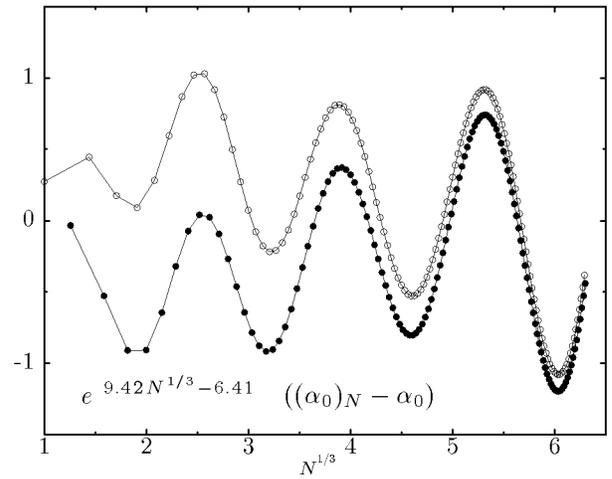

FIG. 2. Oscillatory behavior around the exponential approach to the limiting value of $\alpha_0$

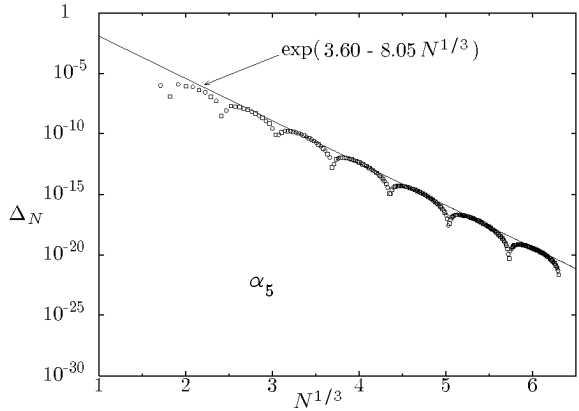

TABLE II. Ground-state energies from strong-coupling series expansion. The lines labeled "lb" and "ub" are the lower and upper bounds of Ref. [6].

| $g/4$ | $n_{\max}$ | $E_0$ |
|---|---|---|
| 0.1 | 15 | 0.559 146 597 503 562 187 0 |
|  | 20 | 0.559 146 201 201 805 544 6 |
|  | 22 | 0.559 146 344 373 873 126 9 |
|  | lb | 0.559 146 327 183 519 576 3 |
|  | ub | 0.559 146 327 183 519 576 7 |
| 0.3 | 15 | 0.637 991 783 178 536 025 3 |
|  | 20 | 0.637 991 783 171 236 149 3 |
|  | 22 | 0.637 991 783 171 280 381 8 |
|  | lb | 0.637 991 783 171 278 528 3 |
|  | ub | 0.637 991 783 171 278 529 6 |
| 0.5 | 15 | 0.696 175 820 765 191 516 9 |
|  | 20 | 0.696 175 820 765 145 887 5 |
|  | 22 | 0.696 175 820 765 145 928 8 |
|  | lp | 0.696 175 820 765 145 925 1 |
|  | up | 0.696 175 820 765 145 928 5 |
| 1.0 | 15 | 0.803 770 651 234 273 812 047 6 |
|  | 20 | 0.803 770 651 234 273 769 350 9 |
|  | 22 | 0.803 770 651 234 273 769 354 1 |
|  | lb | 0.803 770 651 234 273 756 |
|  | ub | 0.803 770 651 234 273 786 |
| 2.0 | 15 | 0.951 568 472 729 500 011 184 213 69 |
|  | 20 | 0.951 568 472 729 500 011 146 930 27 |
|  | 22 | 0.951 568 472 729 500 011 146 930 52 |
|  | lb | 0.951 568 472 729 499 9 |
|  | ub | 0.951 568 472 729 500 1 |

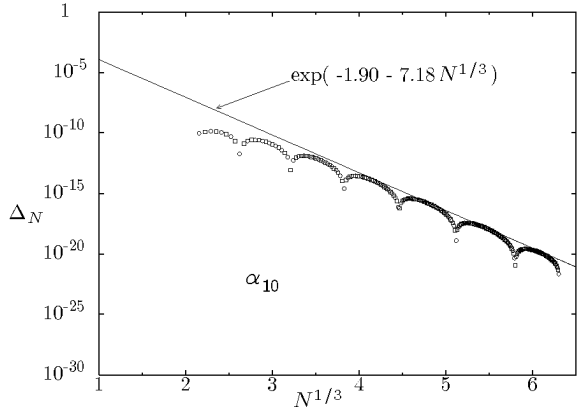